\documentclass{article}
\usepackage{graphicx} 
\usepackage{xcolor}
\usepackage{subcaption}
\usepackage[backend=bibtex,sorting=none]{biblatex} 
\usepackage{geometry}
\usepackage[percent]{overpic}
\usepackage{amsmath} 
\usepackage{authblk} 
\usepackage{hyperref}     
\usepackage{xr-hyper}
\usepackage{xr} 
\usepackage[export]{adjustbox}
\usepackage[normalem]{ulem}

\usepackage{cleveref}

\title{Coherent oscillations in weakly anharmonic NbSe$_2$ qubit }
\author[1]{A. D'Elia}
\author[2,1]{F. Chiarello}
\author[1]{D. Di Gioacchino}
\author[1]{A. S. Piedjou Komnang}
\author[3,4,5]{A.Giachero}
\author[1]{C.Ligi}
\author[1]{G.Maccarrone}
\author[2,1]{F.Mattioli}
\author[6]{C.Pira}
\author[1]{A. Rettaroli}
\author[7,1]{J. Rezvani}
\author[1]{S.Tocci}
\author[1]{C. Gatti}

\affil[1]{INFN Laboratori Nazionali di Frascati, via Enrico Fermi 54, 00044, Frascati (RM), Italy}
\affil[2]{Istituto di Fotonica e Nanotecnologie CNR, 00156 Roma, Italy}
\affil[3]{Dipartimento di Fisica, Università di Milano-Bicocca, Piazza Della Scienza, 3, 20126, Milano, Italy}
\affil[4]{INFN Milano Bicocca, Piazza Della Scienza, 3,  20126, Milano, Italy}
\affil[5]{Bicocca Quantum Technologies (BiQuTe) Centre,  Piazza Della Scienza, 3,  20126, Milano, Italy}
\affil[6]{INFN, Laboratori Nazionali di Legnaro, Legnaro, Padova, Italy}
\affil[7]{Sezione di Fisica, scuola di scienze e tecnologie, Università di Camerino, Camerino, Italy}

\date{\today}

\begin{document}


\maketitle

\begin{abstract}
The functionalization of quantum devices to increase their performance and extend their fields of application is an extremely active research area. One of the  most promising approaches is to  replace aluminum with more performant materials. Within this context, van der Waals (vdW) materials are ideal candidates since they would allow to embed their unique properties into qubits. However, the realization of qubits based on vdW materials other than graphene is yet to be achieved. In this work we present a weakly anharmonic NbSe$_2$ qubit. Our device exhibits a relaxation time $T_1=6.5\pm 0.4\ \mu$s  which is roughly 2 orders of magnitude larger of other vdW qubits in addition to robustness to photon noise up to 5-10 thermal photons.  Our work serves as a demonstrator of the advantage of integration of vdW materials into quantum technologies as well as  serving as the first step toward the application of quantum non demolition photon detection protocols in the challenging field of dark matter search.

\end{abstract}
\section{Introduction}

Quantum technologies are awaited as the next  groundbreaking technological revolution that promises to enable futuristic applications like quantum computers, secure data transmission, detectors sensitive to single energy quanta and so on. The building blocks of quantum technology are qubits, in particular superconducting qubits. These kind of devices use Josephson Junctions (JJ) as non linear oscillators to realize quantum two-level systems that can be manipulated  using microwave pulses~\cite{blais2021circuit}. This approach led to devices with coherence times up to hundreds of $\mu s$~\cite{tuokkola2024methods}. Standard JJ are fabricated in aluminum which offers a mature fabrication technology.  However, this approach has some limitations that hinders Al qubit potential like structural defects~\cite{gao2025effects}, the amorphous and inhomogeneous nature of the AlO$_x$ insulating layer~\cite{zeng2015direct}, dielectric losses~\cite{gunnarsson2013dielectric} and creations of spurious two levels system~\cite{dubois2013delocalized,faoro2006quantum}. All these factors contribute to the reduction of coherence time and are quite hard to tackle. In principle, these limitations can be mitigated by using van der Waals (vdW) materials as a platform to fabricate JJs. This class of materials is characterized by  high crystallinity and atomic clean interfaces that can potentially improve qubits coherence properties. Most importantly, vdW materials when exfoliated down to a thickness of few atomic layers, exhibit novel features that will be inherited by the JJs introducing new functional properties in quantum devices. This would expand the fields of applications where quantum technologies can be relevant~\cite{liu20192d}, improve qubits density per chip~\cite{antony2021miniaturizing,wang2022hexagonal} as well as introduce new qubit design~\cite{brosco2024superconducting}. The process of integration of vdW materials into quantum technologies has already begun leading to the realization of unique devices like graphene based gatemon~\cite{wang2019coherent}, infra red single photon detector~\cite{di2024infrared}, NbSe$_2$-graphene hybrid SQUID~\cite{zalic2023high} and  superconducting diodes~\cite{wu2022field, bauriedl2022supercurrent}.   
The push to realize novel vdW quantum devices with new functional properties finds a natural outcome in the demanding field of dark matter searches where the use of quantum devices is quickly becoming a standard approach~\cite{sushkov2023quantum}. For example, the extreme sensitivity required for dark-photon or axion searches can only be matched by quantum non-demolition photon-detection schemes based on qubit-photon entanglement~\cite{kono2018quantum, dixit2021searching}.
Axion search poses an additional challenge due to the presence of an intense magnetic field that triggers the axion to photon conversion but would disrupt the qubit superconductivity making Al based technology unsuitable for this task. 
A magnetic field resistant qubit would be a game changer enabling the possibility to adopt quantum non demolition schemes as well as searches through direct qubit excitation ~\cite{chen2024search}. In this context, NbSe$_2$ emerges as a natural candidate to fabricate magnetic field resistant quantum devices because it can withstand fields up to 30~T~\cite{xi2016ising}. However, a NbSe$_2$ based qubit hasn't been realized yet. We present the characterization ofa weakly anharmonic NbSe$_2$ qubit that exhibits coherent oscillations with a relaxation time $T_1= 6.5 \ \pm \ 0.4 \ \mu$s and photon noise resistance up 5-10 thermal photons. To our knowledge, this is the first device of its kind. Our result serve as a demonstration of the advantage of integration of vdW materials into quantum devices as well as a first step toward enabling demanding applications like single photon quantum non demolition detection in presence of magnetic field. 

\section{Device fabrication and experimental setup}\label{experimental}
The qubit is composed of a NbSe$_2$-NbSe$_2$ (HQ graphene) omojuction ~\cite{yabuki2016supercurrent} placed inside an Al rectangular cavity. To fabricate the JJ, we used polymer-based dry  transfer techniques~\cite{castellanos2014deterministic}. We assembled the omojunction at INFN-LNF overlapping two different NbSe$_2$ flakes creating a superposition area of $88\ \mu\mbox{m}^2$. From optical contrast, we estimated the flakes thickness to be about $10-15$~nm. The JJ was shunt by two circular Al pads of diameter 1000~$\mu$m and separated by 1483~$\mu$m. The pads were fabricated at CNR-IFN with electron beam lithography and lift off techniques. To enhance the lift off quality, a bilayer resist structure was used. It was made of a first layer of PMMA/MA (AR-P 617.08) followed by  a PMMA 6\% (AR-P 669.06) layer. The bilayer was covered by means of an anti-charging film (AR-PC 5090). A Raith Voyager 50~kV EBL system was used to expose the pattern made for the two circular Al pads and two 2~$\mu$m wide wires that connect them to the central area where the junction was then fabricated. After development (MIBK:IPA 1:3), 80~nm of aluminum are e-gun evaporated and hence lifted off using hot acetone. An optical image of the NbSe$_2$ JJ is shown in Fig.~\ref{fig:optical}. The aluminum cavity, of purity 99.999\% and dimensions $\mbox{L}_x\times\mbox{L}_y\times\mbox{L}_z=26\times36\times8~\mbox{mm}^3$, was produced at INFN-LNL. Two treatments were done to lower the surface roughness after mechanical machining: a 24 hours tumbling treatment with coco powders and a subsequent 1 hour electro-polishing in solution H$_3$PO$_4$:BuOAc 65:35 at temperature 60~$^{\circ}$C. The two processes removed about $20-30~\mu$m of Al from the surface.

\begin{figure}[h!]
    \centering
    \includegraphics[width=0.5\textwidth]{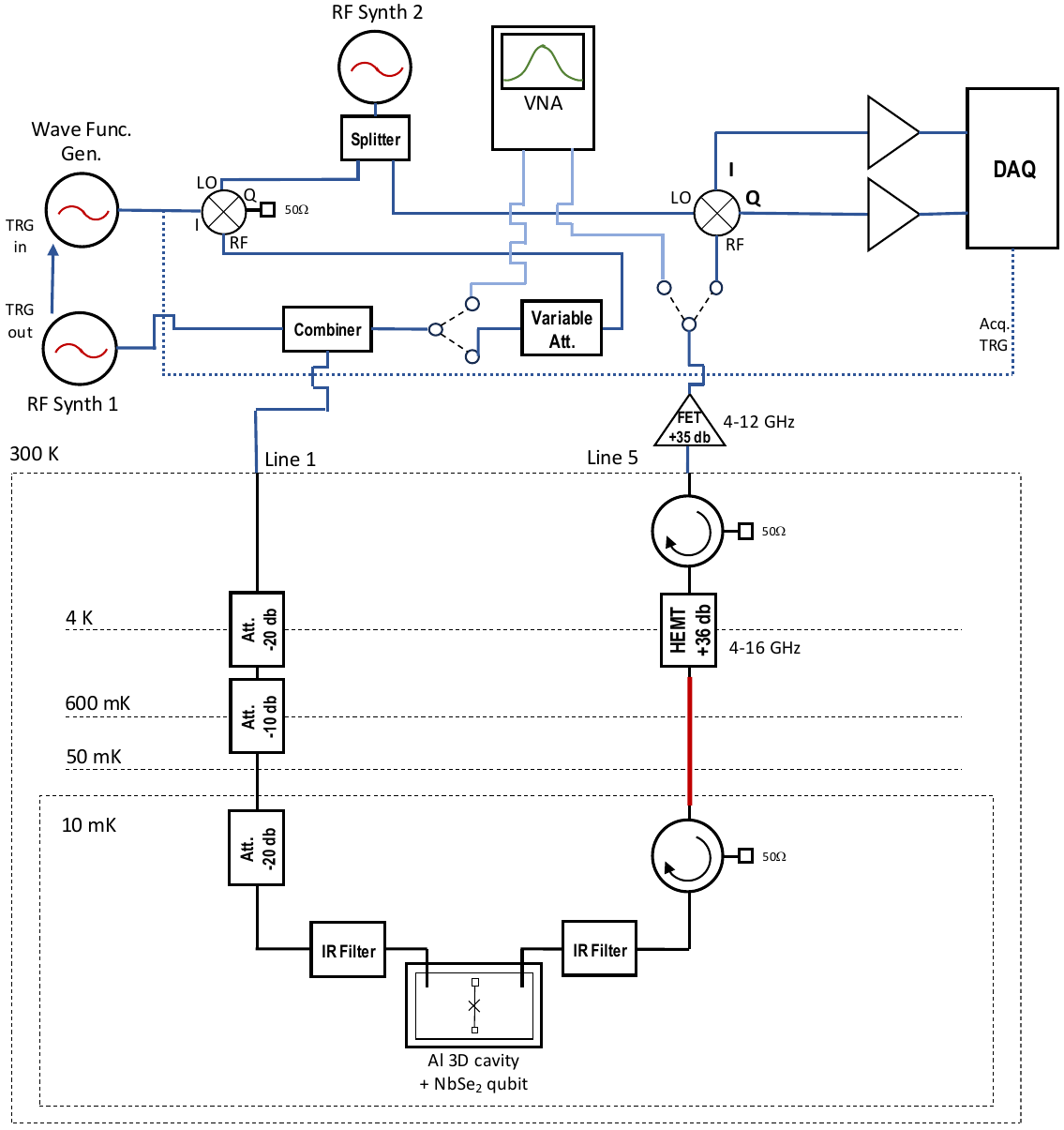} 
    \caption{Experimental scheme. The upper part shows the room temperature instrumentation and wiring, the lower part in the 300~K dashed black box represents the setup in the dilution refrigerator. The red line is a superconducting NbTi cable for low-loss transmission. The I--Q mixers serving as vector-modulator and down-converter are of the same model. The small dashed lines with open circles at their edges represent mechanical switches. More details in the main text.}
    \label{fig:schemaRF}
\end{figure}

The device was measured at a temperature of $\sim 20$~mK in the dilution refrigerator of the COLD (cryogenic laboratory for detectors) laboratory at INFN-LNF. The measurement setup is shown in Fig.~\ref{fig:schemaRF}. The dashed black lines and boxes indicate the different temperature stages of the cryostat. Line~1 refers to the attenuated input RF transmission line, whereas Line~5 refers to the low-loss, amplified output line. The Aluminum cavity was protected from IR radiation with IR filters while the output was isolated from 4~K backward radiation coming from the HEMT (High Electron Mobility Transistor) amplifier by an isolator. The upper part of the diagram shows the instrumentation and wiring used to characterize the device. Switches allowed to choose between spectroscopic and time-domain measurements. The former was conducted using a Vector Network Analyzer (VNA) while the latter via the vector modulation of a harmonic signal (mixer on the left) followed by down-conversion before digitization (mixer on the right). The signals from the VNA or vector modulation and from an RF synthesizer (RF~Synth.~1) were routed via the combiner to Line~1. The NbSe$_2$ qubit was controlled by sending microwave pulses of frequency 12.611~GHz whereas its state was determined by cavity transmission measurements obtained by sending RF pulses of frequency 7.1873~GHz, with estimated attenuation along the input line of -85 and -75~dB, respectively.  The output signal was down-converted to base-band I-Q waveforms which were further amplified and acquired by a digital oscilloscope (DAQ). This system enabled the qubit control, pulse calibration through Rabi experiments and relaxation time measurements.

 We observed cavity resonant modes at the frequencies 7.1873, 10.40, 13.45, 17.85~GHz, and identified them as the modes $TM_{110}$, $TM_{120}$, $TM_{130}$ and $TM_{140}$. We used the mode $TM_{110}$ as readout mode. It was less coupled to the qubit than the mode at 13.64 GHz (See Appendix~\ref{sec:cavitysimul}) but it was inside our amplification bandwidth ($4-12$~GHz).

\section{Results}

\begin{figure}
    \centering
    \hfill
    \begin{subfigure}[b]{0.45\textwidth}
    \includegraphics[width=6.5cm, height=6cm]{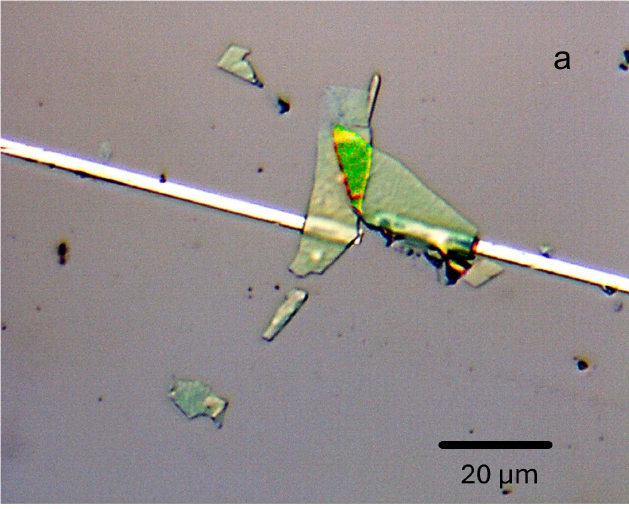}
    \phantomcaption    
    \label{fig:optical}
    \end{subfigure}
    \hfill
    \begin{subfigure}[b]{0.45\textwidth}

        \includegraphics[width=\textwidth]{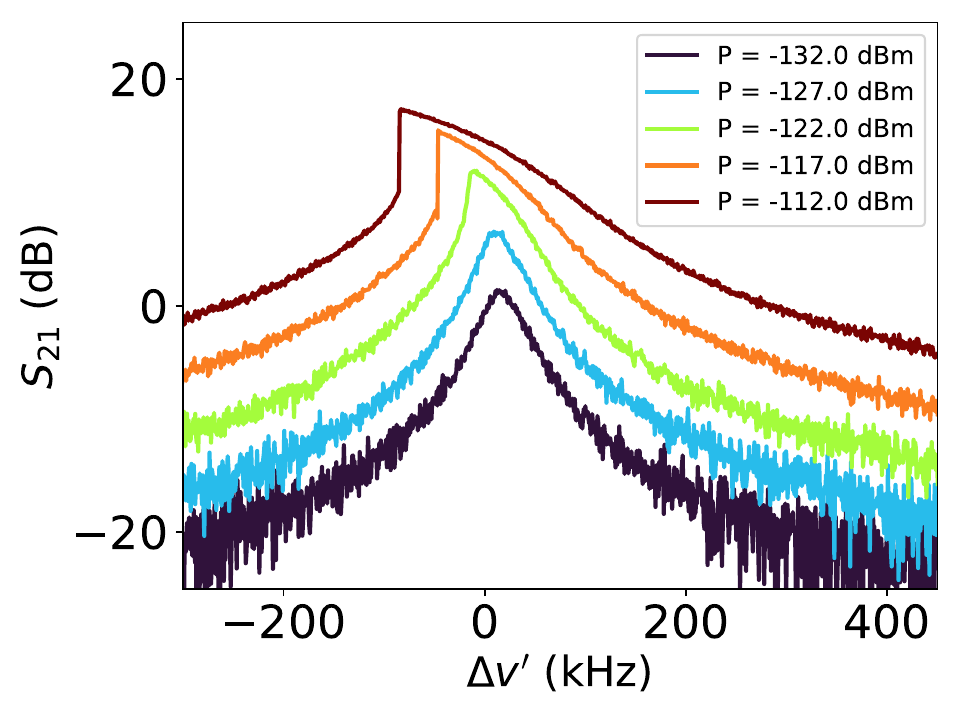}
        \phantomcaption    
        \label{fig:S21}
    \end{subfigure}
    \hfill

    \begin{subfigure}[b]{0.48\textwidth}
        \includegraphics[width=\textwidth]{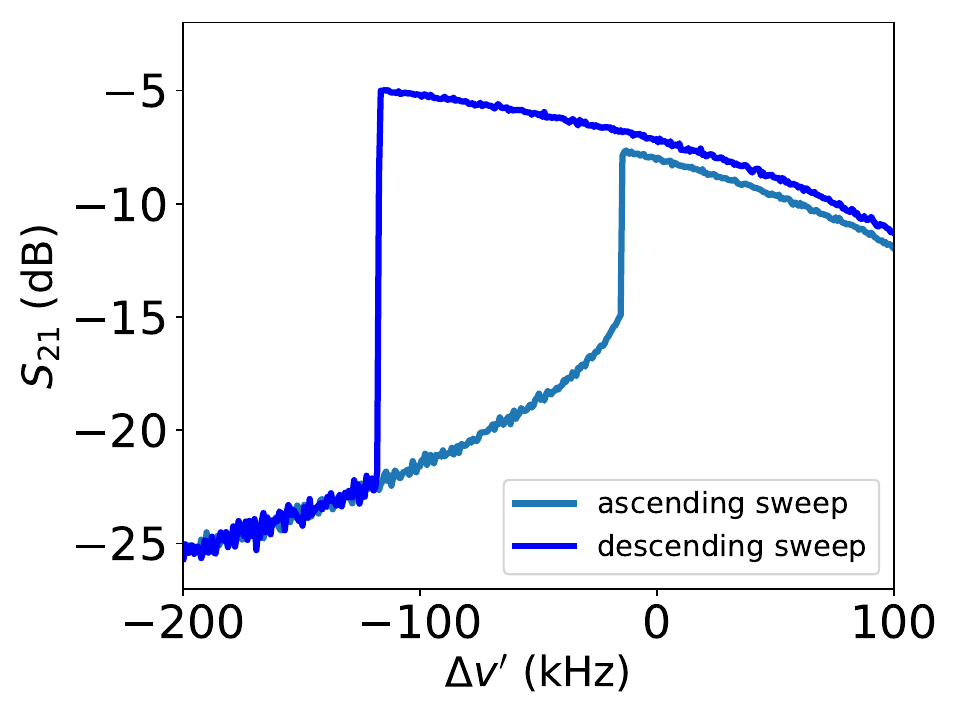}
        \phantomcaption    
    \label{fig:isteresi}
    \end{subfigure}
    \begin{subfigure}[b]{0.48\textwidth}
        \includegraphics[width=\textwidth]{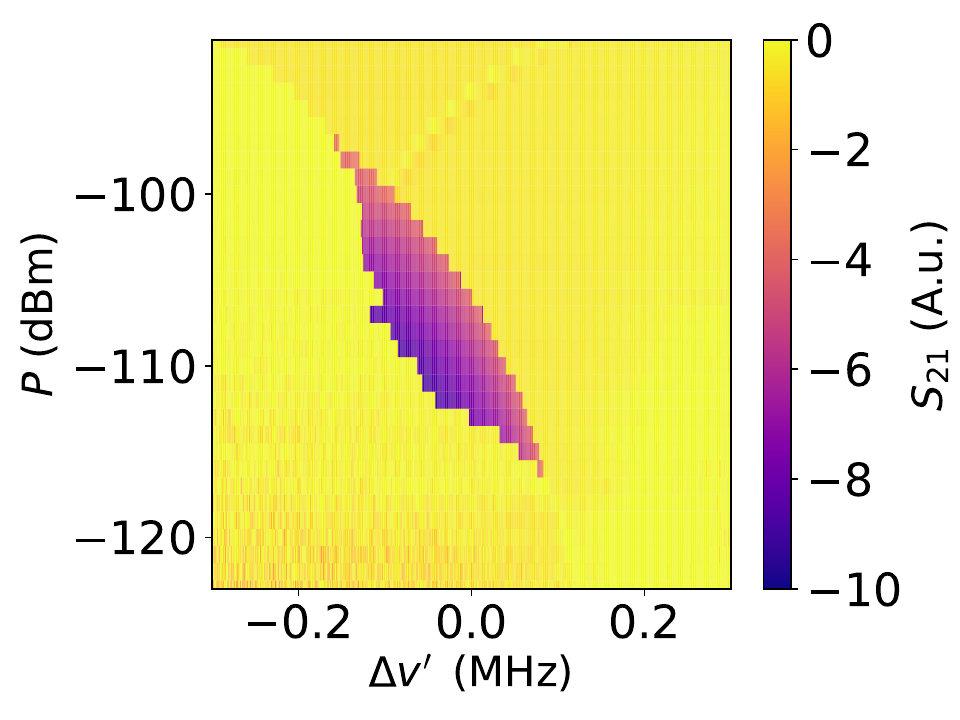}
        \phantomcaption    
    \label{fig:isteresi_mappa}
    \end{subfigure}
    \caption{a): Optical image of the NbSe$_2$-NbSe$_2$ JJ. The overlap area (88 $\mu\mbox{m}^2$) is reported in false color for visibility. b) Dressed cavity transmission evolution as a function of the readout power (Read out frequency expressed as difference with $\nu_{r}=7.1873$~GHz). The appearance of a step-like resonance is a fingerprint that our system behaves as a Duffing oscillator. The spectra are vertically shifted for better visualization. c): Hysteresis of the Duffing oscillator, depending if the frequency sweep spans from low to high frequency value (ascending) or viceversa (descending). The transmission curves are acquired with  readout power at resonator input, P$= -114$~dBm (the x axis is expressed as detuning from $\nu_{r}=7.18723$~GHz). d): 2D map displaying the Duffing oscillator hysteresis where we plotted the difference of cavity transmission measured from ascending and descending sweeps. The x axis is expressed as detuning from $\nu_r=7.187$ MHz  }
    \hfill

    \label{pinne}
\end{figure}

We initially characterized the vdW device by performing the spectroscopy of the cavity mode $TM_{110}$. Increasing the readout power, the cavity transmission showed a discontinuous shape typical of Duffing oscillators (See Fig.~\ref{fig:S21})~\cite{agarwal2018influence,murch2011quantum}. We did not observe cavity frequency switch to the bare value for readout powers up to $-92$~dBm at the cavity input~\cite{reed2010high}. This indicates a high photon noise resistance of our device that demonstrated to work also in presence of 5-10 thermal photons (See Appendix~\ref{psn}). 
Such resistance is attributable to the high T$_c$ ($\sim 7$~K for bulk NbSe$_2$) of the superconducting electrodes as well as the high $I_c$ of the JJ that allowed our device to withstand higher intracavity fields with respect to standard materials. The onset from a symmetrical Lorentzian shape to a step-like resonance was around $-120$~dBm. The bending toward lower frequencies indicates a negative non-linearity (softening oscillator)~\cite{wawrzynski2022origin}.  The non-linear nature also manifested in hysteretic behavior depending whether the readout frequency was swept ascending (from low to high frequency) or descending (from high to low frequency) as reported in Fig.~\ref{fig:isteresi} and~\ref{fig:isteresi_mappa}~\cite{wawrzynski2022origin,guo2010quantum,bishop2010response}.

\begin{figure}
    \centering
   
    \begin{subfigure}[b]{0.48\textwidth}
        \includegraphics[width=\textwidth]{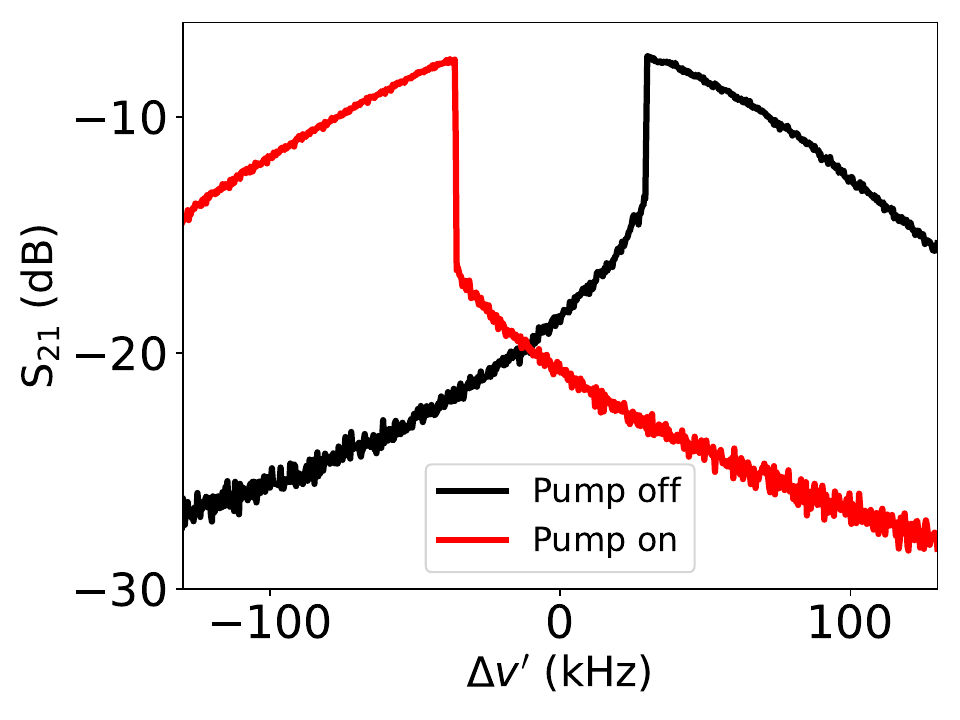}
        \phantomcaption    
        \label{fig:pinne}
    \end{subfigure}
    \begin{subfigure}[b]{0.48\textwidth}
        \includegraphics[width=\textwidth]{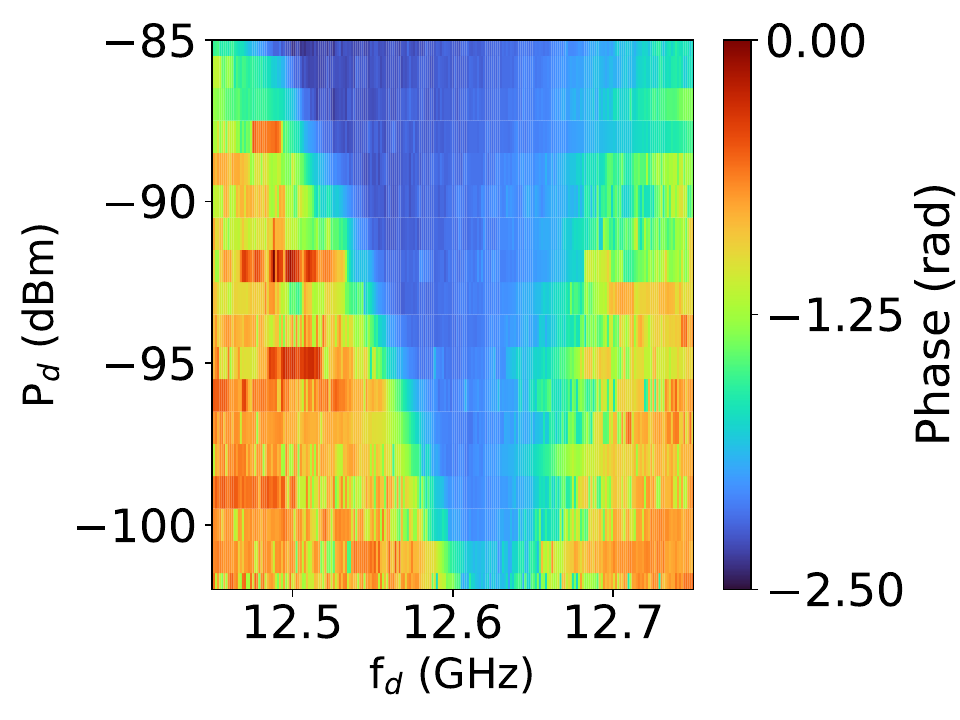}
        \phantomcaption    
        \label{fig:2tones}
    \end{subfigure}
    
    \begin{subfigure}[b]{0.48\textwidth}
        \includegraphics[width=\textwidth]{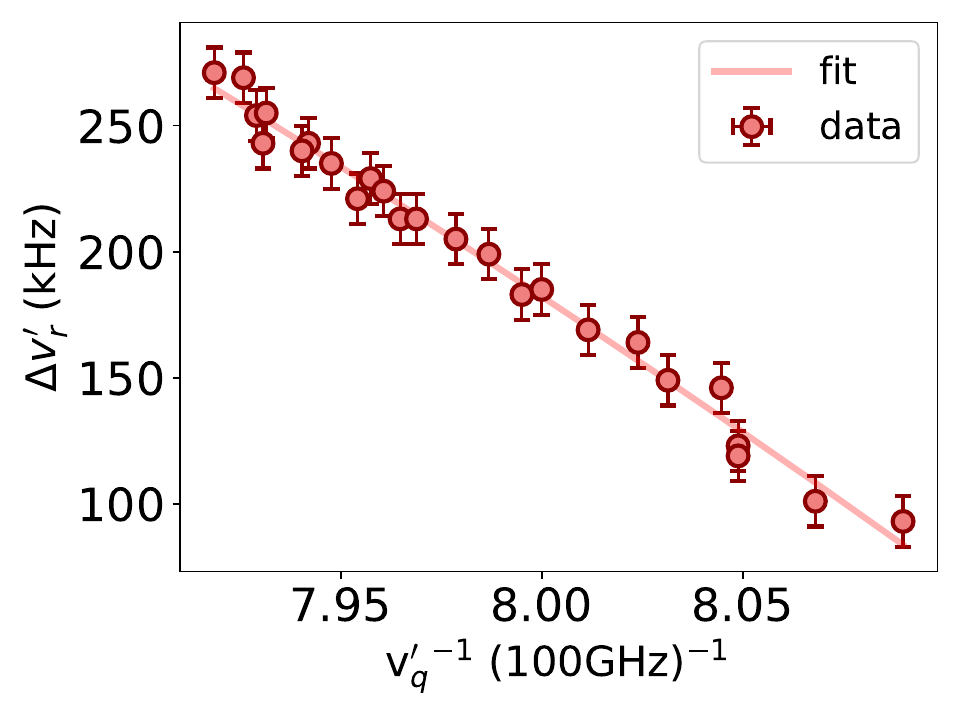} 
        \phantomcaption    
        \label{fig:vwvsvr}
    \end{subfigure}
    \begin{subfigure}{0.48\textwidth}
        \includegraphics[width=\textwidth]{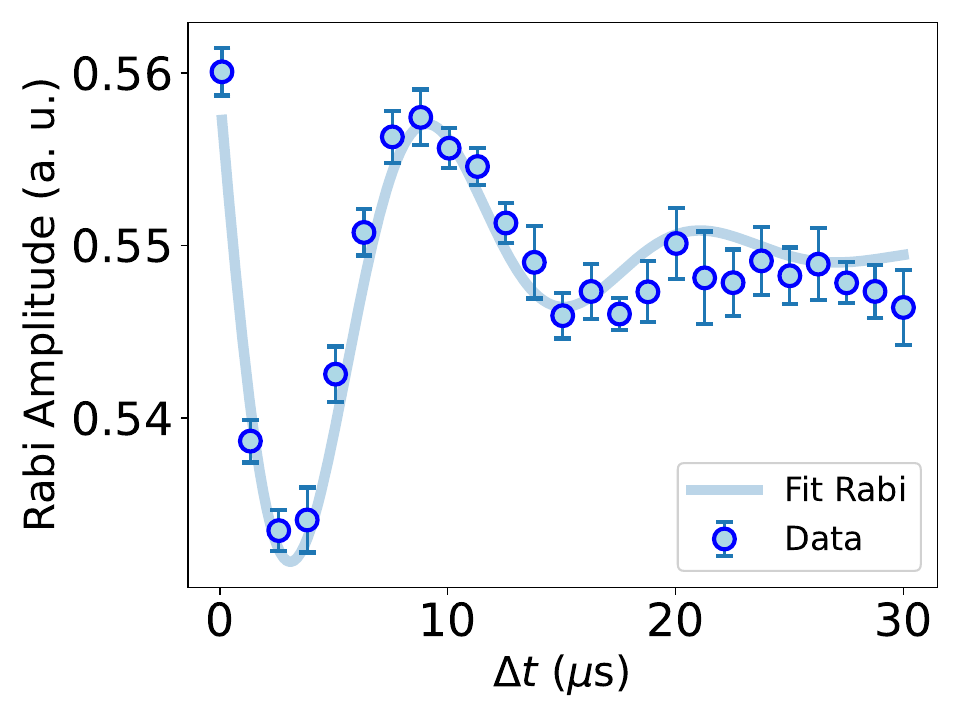}
        \phantomcaption    
        \label{fig:rabi}
    \end{subfigure}
    \caption{a): Switch from softening (pump off) to hardening (pump on: $-75$~dBm) Duffing oscillator (the $x$ axis is expressed as detuning from $\nu_{c}=7.1873$~GHz).b): Two-tone spectroscopy of the device as a function of the drive tone frequency and power.  The readout tone is set at $\nu_{ro}$ =7.1873~GHz with power P$_{ro}= -132$~dBm. c): Dressed resonator frequency $\Delta\nu_{r}'$ (expressed as detuning from $\nu_c=7.187$~GHz) as a function of the inverse qubit frequency $\nu^{-1}_q$. d) Rabi oscillation of the qubit for a drive power P$=-95$~dBm as a function of the time length of the driving pulse $\Delta t$.}
    \label{fig:2 tones spectroscopy}
\end{figure}

Given the small anharmonicity, we described the device as
a multi level qubit dispersively coupled to a resonator. The dressed cavity Hamiltonian is ($\hbar=1$) \cite{blais2021circuit}:
\begin{equation}
    H = \left(\omega_r+\chi b^{\dagger}b\right)a^{\dagger}a+
    \frac{K_a}{2} a^{\dagger 2}a^2+\omega_qb^{\dagger}b+\frac{K_b}{2} b^{\dagger 2}b^2 ,
    \label{hamiltonian}
\end{equation}
where, $a$ ($a^{\dagger}$) is the annihilation (creation) operator of a cavity mode, $b$ ($b^{\dagger}$) the qubit annihilation (creation) operator and $\omega_r$ and $\omega_q/2\pi=12.611$~GHz are the cavity mode and qubit frequencies. The Kerr terms $K_{a,b}$ and $\chi$ are:
\begin{center}
\begin{align}
K_a=\alpha\left(\frac{g}{\Delta}\right)^4 
\label{selfcavity}\\
K_b=\alpha=-E_c
\label{selfqubit}\\
\chi\simeq\frac{-2g^2\alpha}{\Delta(\Delta-\alpha)}\,.
\label{dispshift}
\end{align}
\end{center}
$K_{a,b}$ are the self-Kerr terms of the cavity and qubit, respectively, $\chi$ is the cross-Kerr term, $E_c$ the charging energy, $g$ the cavity qubit coupling, $\alpha$ the anharmonicity and  $\Delta=\omega_q-\omega_r$ the detuning. The sign of the self-Kerr term $K_a$ is negative coherently with the observation of a softening Duffing oscillator. However, driving the qubit above a power threshold P$\approx -100$~dBm at the cavity input, the curvature of the dressed cavity transmission switched to an hardening Duffing oscillator as shown in Fig.~\ref{fig:pinne} implying a sign swap of the Kerr term.  This is due to the contribution of higher order terms in the expansion of the JJ potential (See Appendix~\ref{pinneswap}) and is non-perturbatively derived in the limit of a 2-level system~\cite{mavrogordatos2017simultaneous}.\\

We continued the device characterization with the qubit spectroscopy performed by sending a driving tone with the frequency of the qubit to the device while reading the phase of the signal transmitted by the cavity mode TM$_{110}$ and repeating the measurement for different frequencies and strengths of the driving tone. The resulting map of this two-tone spectroscopy, shown in Fig.~\ref{fig:2tones}, has the continuous feature characteristic of small-anharmonicity systems. We estimated the anharmonicity of our device calculating the charging energy of the capacitance of the NbSe$_2$ JJ. The area of the JJ, measured using optical microscopy, is 88~$\mu\mbox{m}^2$. Since flakes were exfoliated in air, a thin layer of NbO$_x$ formed at the interface of the two NbSe$_2$ flakes. The thickness is assumed to be $\sim 1$~nm~\cite{kim2020non,zhang2020nonreciprocal} considering that the total exposure time to air for the whole fabrication process was less than 20 minutes. With these parameters, we calculated the geometrical capacitance $C \sim 15$~pF  ($\epsilon_r\sim 20$~\cite{herzig2019improvement}) and the anharmonicity $\alpha/2\pi \sim -1.3$~MHz, that combined with the value of the qubit frequency lead to an estimate of the JJ critical current $I_c=31\ \mu$A, a value consistent to those appearing in other studies on NbSe$_2$ omojunctions~\cite{sunny2025overdamped}.

We estimated the dipole coupling $g_{110}$ of the qubit to the cavity mode $TM_{110}$ by measuring the dressed-cavity frequency while driving the qubit at different powers (details in Appendix~\ref{cavityvsdrivepower}). The dressed-cavity frequency is expected to change according to
\begin{equation}
\label{vr vs vq-1}
\omega^{\prime}_r=\omega_r+\chi b^{\dagger}b=\omega_r-\frac{4g_{110}^2}{\Delta(\Delta-\alpha)}\frac{\alpha}{2}b^{\dagger}b
\end{equation}
We used the readout power P$=-121$~dBm at the cavity port while varying the qubit drive-power in the range $(-104;-69)$~dBm  similar to the one used for the two-tone spectroscopy (See Fig.~\ref{fig:2tones}). To reduce the uncertainty due to power calibration and knowledge of the anharmonicity $\alpha$, we determined the term $\alpha b^{\dagger}b$ by measuring the shift of the qubit frequency as a function of the drive power (See Fig.~\ref{fig:2tones})
\begin{equation}
\label{vr vs vq-2}
\omega_q^{\prime}=\omega_q+\frac{\alpha}{2}b^{\dagger}b
\end{equation}
This allowed us to plot the dressed cavity frequency as a function of the qubit frequency (See Fig.~\ref{fig:vwvsvr}) and derive their relation
\begin{equation}
\label{vr vs vq-3}
\omega^{\prime}_r=\omega_r-\frac{4g_{110}}{\Delta(\Delta-\alpha)}(\omega^{\prime}_q-\omega_q)\,.
\end{equation}
We used Eq.~\ref{vr vs vq-3} to fit the data, estimating $g_{110}/2\pi =67\ \pm17$~MHz for $\alpha/2\pi=-1.3$~MHz. This value is in good agreement with the value $g_{110}^{EPR}/2\pi =54$~MHz extracted from electromagnetic simulations with the EPR~\cite{pyEPR} method (See Appendix~\ref{sec:cavitysimul}).

\begin{figure}
    \centering
    \begin{subfigure}{0.45\textwidth}
        \includegraphics[width=\textwidth]{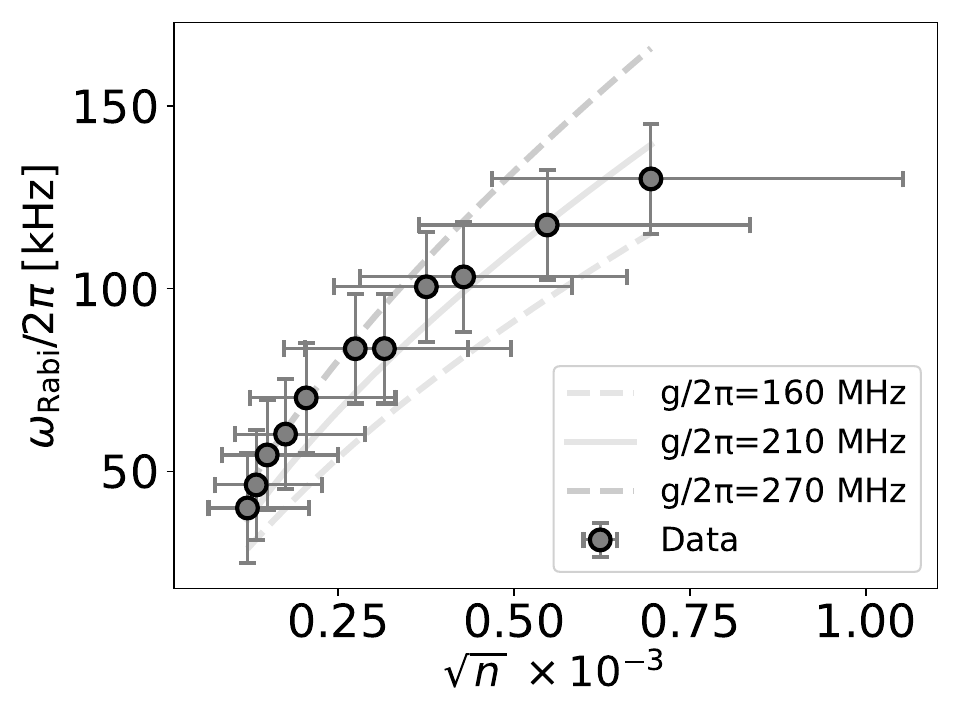} 
        \phantomcaption    
        \label{fig:rabivspower}
    \end{subfigure}
        \begin{subfigure}{0.45\textwidth}
        \includegraphics[width=\textwidth]{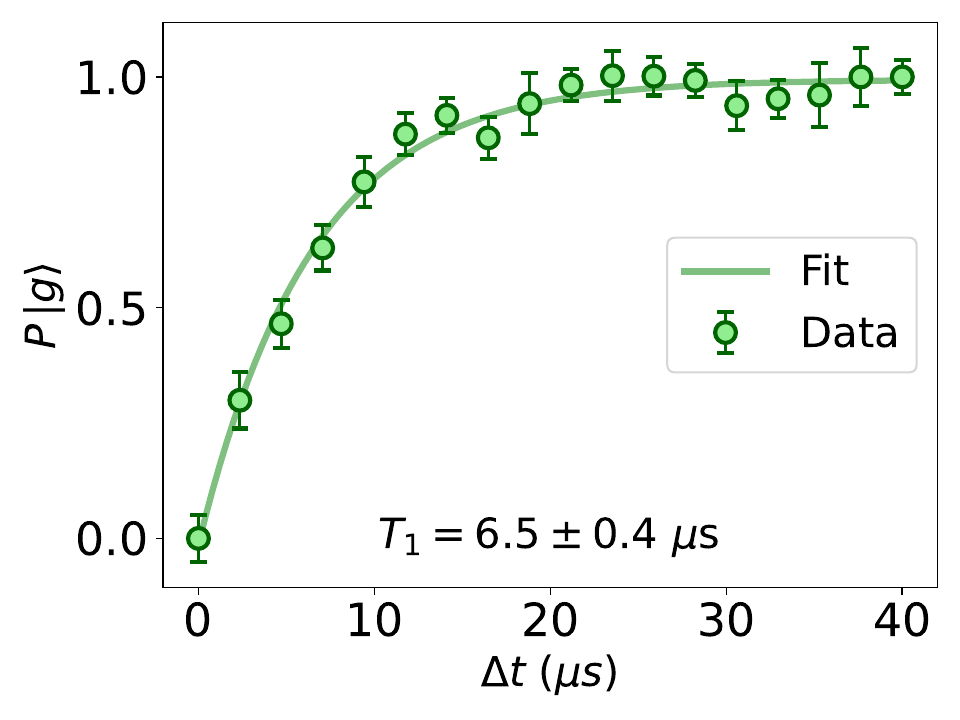} 
        \phantomcaption    
        \label{fig:T1}
    \end{subfigure}
    \caption{ a) Rabi frequency as a function of the drive amplitude expressed as squared root of the effective cavity occupation-number. The deviation from the linear behavior points to a multi-level process. Error bars are dominated by systematics and the horizontal ones are correlated. The data are reported along with the simulated curve obtained solving the time evolution of eq. \ref{simulatio_hamil} in appendix \ref{cavityattenuationII} for different values of $g/2\pi$.  b) Decay time $T_1$ expressed as a function of the delay between the $\pi$ excitation pulse and the readout, acquired with power P$=-101$~dBm. The data are expressed as the probability to find the qubit in the ground state.}
    \label{fig:T1rabivspower}
\end{figure}

We observed Rabi oscillations demonstrating the coherence properties of our device. To do this, we varied the time length of excitation pulses with power in the range P$_d=(-101;-89)$~dBm. As shown in Fig.~\ref{fig:rabi}, our device performed clear coherent oscillations. The Rabi frequency $\nu_{Rabi}=\omega_{Rabi}/2\pi$ has a linear trend as a function of the drive amplitude (expressed as the square root of the effective occupation number of the cavity $n$), consistent with qubit oscillations between $|0\rangle $ and $|1\rangle$ (See Fig.~\ref{fig:rabivspower}).  Increasing the drive power, we observed a non linear trend of $\nu_{Rabi}$. The deviation from linearity indicates that higher levels were excited, with the device acting as a multi-level system~\cite{thery2024observation,claudon2008rabi, amin2006rabi}. This is expected by its small anharmonicity.

The data are reproduced simulating the time evolution of a Duffing oscillator with coupling $g_{130}^{sim}/2\pi=210\pm 60$~MHz and anharmonicity $\alpha/2\pi=-0.03$~MHz. The first is in reasonable agreement with our estimate from experimental data, $g_{130}/2\pi=257\pm 60$~MHz, and with the EPR simulation of $g_{130}^{EPR}/2\pi=210$~MHz, while the anharmonicity is smaller than our estimate (See Appendix~\ref{kerrdata} and ~\ref{cavityattenuationII}). We considered the mode $TM_{130}$, the closest in frequency to the qubit and with the highest dipolar coupling. 

We measured the qubit relaxation time $T_1$ by first calibrating the $\pi$-pulse duration from Rabi oscillations. To minimize the leakage to higher levels, we adopted a drive power P$=-101$~dBm, corresponding to the minimum power for which Rabi oscillations are observable. Then, after preparing the qubit in the excited state, we measured the probability to find it in the ground state after a delay time $\Delta t$ (Fig.~\ref{fig:T1}). From this distribution we determined  $T_1= 6.5 \ \pm \ 0.4 \ \mu$s. This result is about two orders of magnitude larger than the previously reported coherence time for graphene-based vdW qubit~\cite{wang2019coherent}. This long relaxation-time is probably due the combined contribution of the high crystalline-quality of the NbSe$_2$, that minimizes lattice defects, and to the high $I_c$ of the junction that grants a robust noise tolerance. 

Ramsey measurements aimed at estimating the decoherence time $T_2$ were not successful due to the long duration of the $\pi$ pulses ($\sim 5\,\mu$s). For long $\pi/2$ pulses, the qubit starts relaxing and dephasing during the same excitation, reducing the probability of observing the interference process.

\section{Discussion}
In conclusion, we have successfully demonstrated the functioning of a weakly anharmonic NbSe$_2$ qubit. Our device exhibits the features of a Duffing oscillator with step-like resonances and hysteresis cycle. We have shown that it is possible to change the sign of the nonlinearity and thus change the behavior of the device from a softening to a hardening oscillator. We observed coherent oscillations with a relaxation time $T_1\ = 6.5 \pm0.4\ \mu$s that is almost 2 orders of magnitude larger than that of devices made with other vdW materials~\cite{wang2019coherent}.  In addition, our device has proven to be very robust against photon noise working in presence of 5-10 thermal photons in the cavity.  This features could ease the use of quantum devices in noisy environments with respect to standard Al based technology. 
To our knowledge, this is the first device of its kind and we believe this work can pave the way for integration of vdW materials into quantum devices thus combining two of the most active field of research of the latest years.  
The integration of vdW materials into quantum technologies holds the potential to unlock novel physical properties and enable measurement schemes and functionalities otherwise inaccessible with conventional platforms. In particular, qubits based on NbSe$_2$ offer a promising pathway toward quantum-enhanced axion detection, both through direct excitation of a magnetic field resistant transmon within axion cavities~\cite{chen2024search} and facilitating quantum non-demolition measurement protocols. In this study, we have demonstrated the fabrication of a homojunction via mechanical exfoliation — a technique especially well-suited for applications requiring a limited number of high-coherence qubits. While the resulting device exhibits a small anharmonicity, this parameter can be finely tuned either through nanometric alignment techniques reducing the effective JJ area, or by leveraging high-resolution lithography, such as electron-beam patterning~\cite{sunny2025overdamped}. These findings establish a foundation for future vdW-based quantum devices tailored for precision sensing and fundamental physics experiments.

\section{Acknowledgments}
This work was supported by RESILIENCE, Qub-IT and QUART\&T, projects funded by the Italian Institute of Nuclear Physics (INFN) within the Technological and Interdisciplinary Research Commission (CSN5), by PNRR MUR projects PE0000023-NQSTI and CN00000013-ICSC. We thank Massimo Bauco (Rohde \& Schwarz Roma) for the support with reverse-sweep of the VNA, and Marco Beatrici for the technical support in the lab.

\printbibliography

\appendix
\section*{Appendix}
\section{Simulation of cavity modes and calculation of their couplings to the qubit}\label{sec:cavitysimul}

\begin{figure}[htbp]
    \centering
    
    \includegraphics[width=0.6\textwidth]{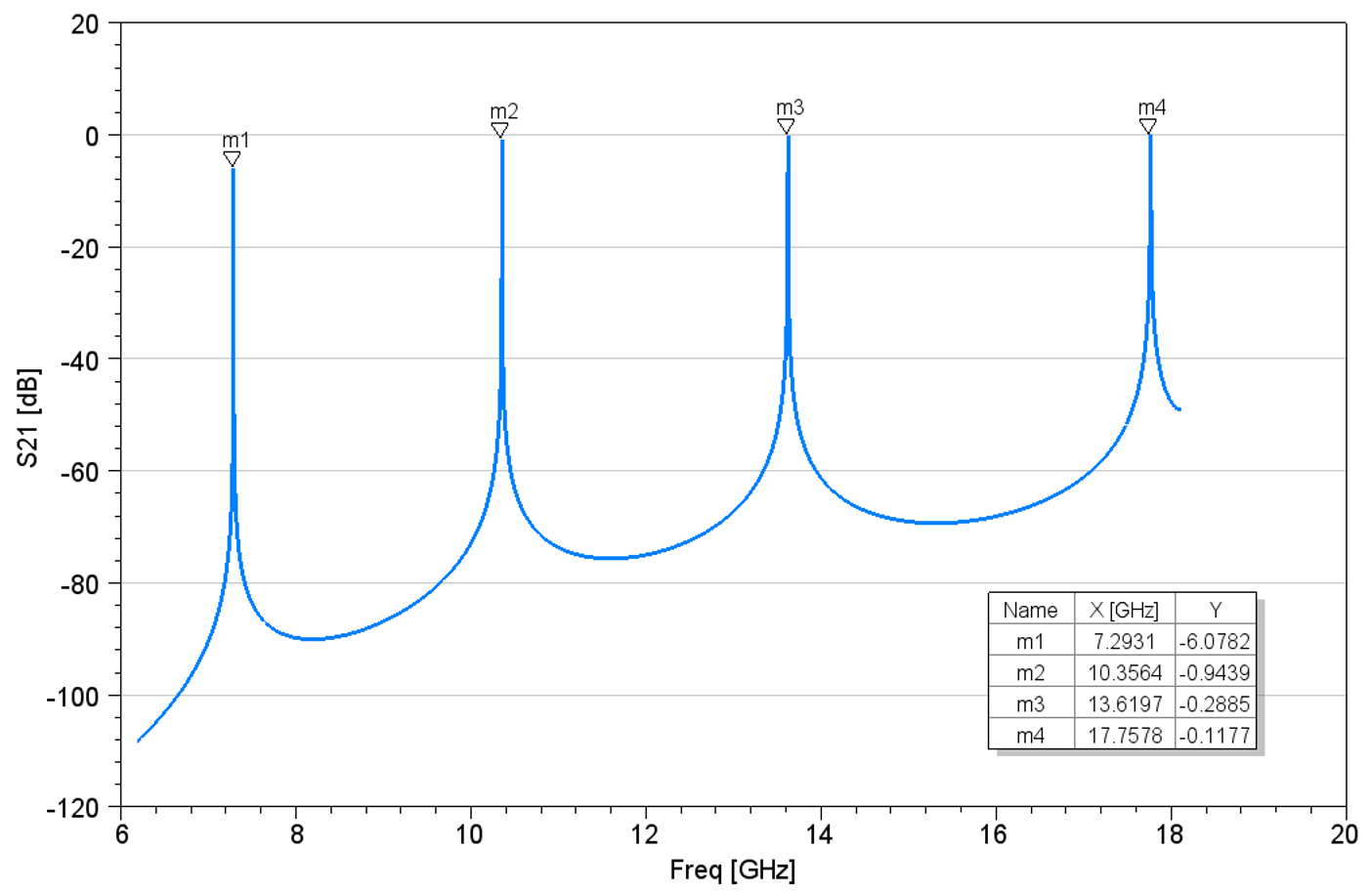}
    \caption{Simulated transmission of the four main modes of the Al cavity.}
    \label{fig:fullwidth}
\end{figure}
\begin{figure}[htbp]
    \centering

    \begin{subfigure}[b]{0.49\textwidth}
        \centering
        \includegraphics[width=\linewidth]{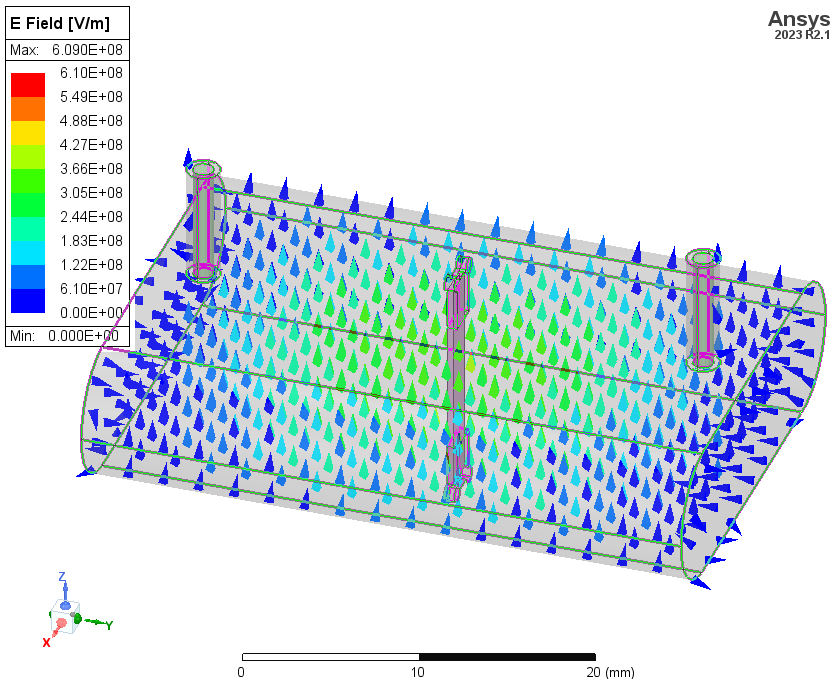}
        \phantomcaption    
        \label{fig:mode1}    
    \end{subfigure}
    \hfill
    \begin{subfigure}[b]{0.49\textwidth}
        \centering
        \includegraphics[width=\linewidth]{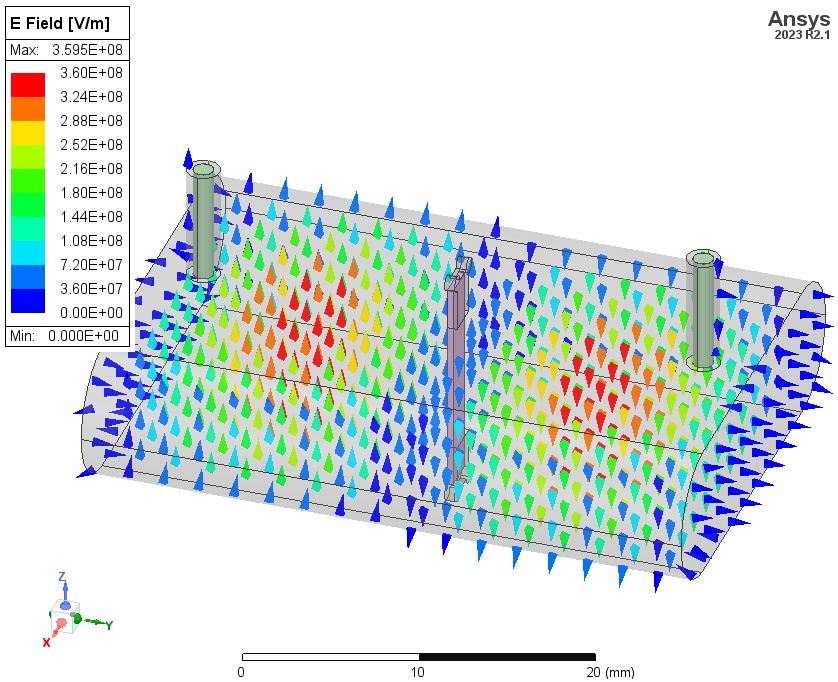}
        \phantomcaption    
        \label{fig:mode2}    
    \end{subfigure}

    \begin{subfigure}[b]{0.49\textwidth}
        \centering
        \includegraphics[width=\linewidth]{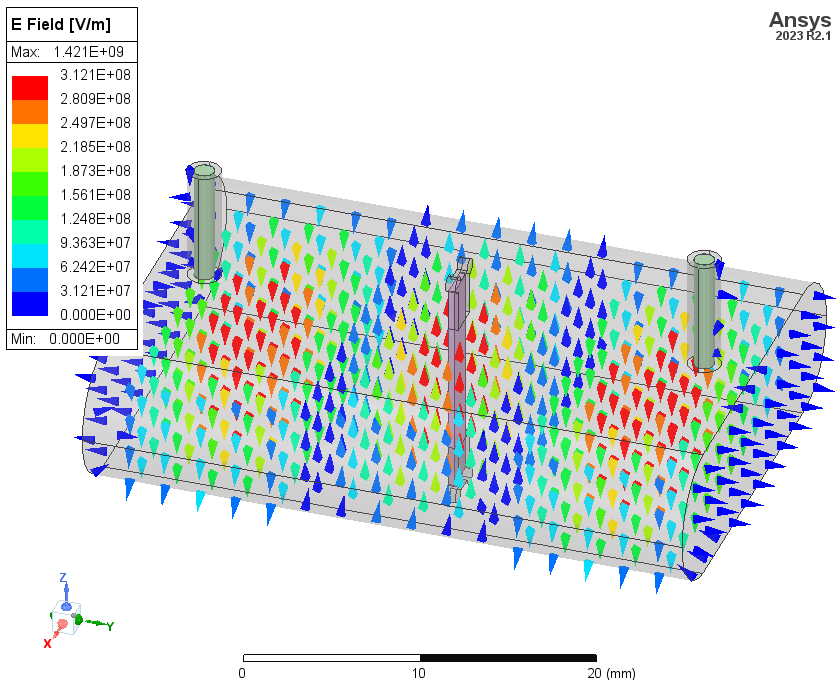}
        \phantomcaption    
        \label{fig:mode3}    
    \end{subfigure}
    \hfill
    \begin{subfigure}[b]{0.49\textwidth}
        \centering
        \includegraphics[width=\linewidth]{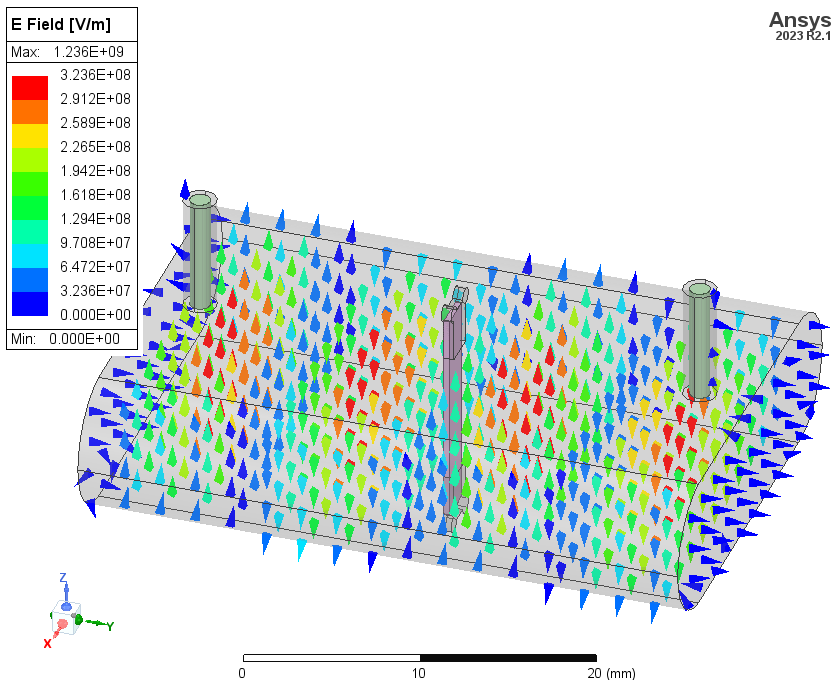}
        \phantomcaption    
        \label{fig:mode4}    
    \end{subfigure}

    \caption{Electric field configuration of the four main cavity modes. Upper left: TM$_{110}$ mode, 7.29~GHz; Upper right: TM$_{120}$ mode, 10.35~GHz; Lower left: TM$_{130}$ mode, 13.35~GHz; Lower right: TM$_{140}$ mode, 17.48~GHz}
    \label{fig:electric field modes}
\end{figure}

We simulated the electromagnetic response of our device with Ansys HFSS~\cite{Ansys:HFSS} and determined couplings and  dispersive shifts with the EPR~\cite{pyEPR} method. The simulated transmission and the electric field distribution of these modes are shown in Fig.~\ref{fig:fullwidth} and~\ref{fig:electric field modes}. In the simulation, the end of the antennas conductors are positioned symmetrically at the entrance of the cavity as in the experimental setup. For the cavity, we obtained the frequencies 7.29, 10.35, 13.35, and 17.48~GHz corresponding to the modes TM$_{110}$ (m1), TM$_{120}$ (m2), TM$_{130}$ (m3), TM$_{140}$ (m4), in good agreement with the measured values 7.187, 10.4, 13.45, and 17.85~GHz, respectively. We obtained 45, 200, and 600~kHz for the widths of modes 1, 3 and 4. We experimentally measured 53~kHz for m1, reasonably in agreement with simulation and compatible with the losses induced by the silicon chip, while we observed a much larger width, about 4~MHz, for m3, probably affected by the non-linearity of the qubit. The simulated values of the widths are very sensitive to changes in the antenna positioning at the level of a few tenths of a millimeter.

While the electric field in the mode m1,3 and 4 is different from zero at the point where the qubit is located (See Fig.~\ref{fig:electric field modes}), leading to a good coupling to the qubit,  mode m2 exhibits a zero of the electric field at the qubit position (See Fig~\ref{fig:mode2}) leading to a vanishing coupling. These considerations were confirmed experimentally. We used simulations and the EPR method to estimate the dipolar couplings between the qubit and the three modes. Setting the JJ parameters as C$_j\ =15.3~\mbox{pF}$ and L$_j\ =1.04\cdot10^{-2}$~nH, we obtained the anharmonicity $\alpha/2\pi=-1.37$~MHz and the couplings $g_{110}^{EPR}/2\pi=52$~MHz, $g_{120}^{EPR}/2\pi=5$~MHz and $g_{130}^{EPR}/2\pi=210$~MHz for the modes m1, m2, m3, respectively. This estimate is reduced by a factor 3 when the value of the junction capacitance is increased by an order of magnitude. 

\section{Estimation of the photon noise resistance of the NbSe$_2$ qubit } \label{psn}

\begin{figure}
    \centering
    \begin{subfigure}{0.5\textwidth}
        \centering
        \includegraphics[width=\textwidth]{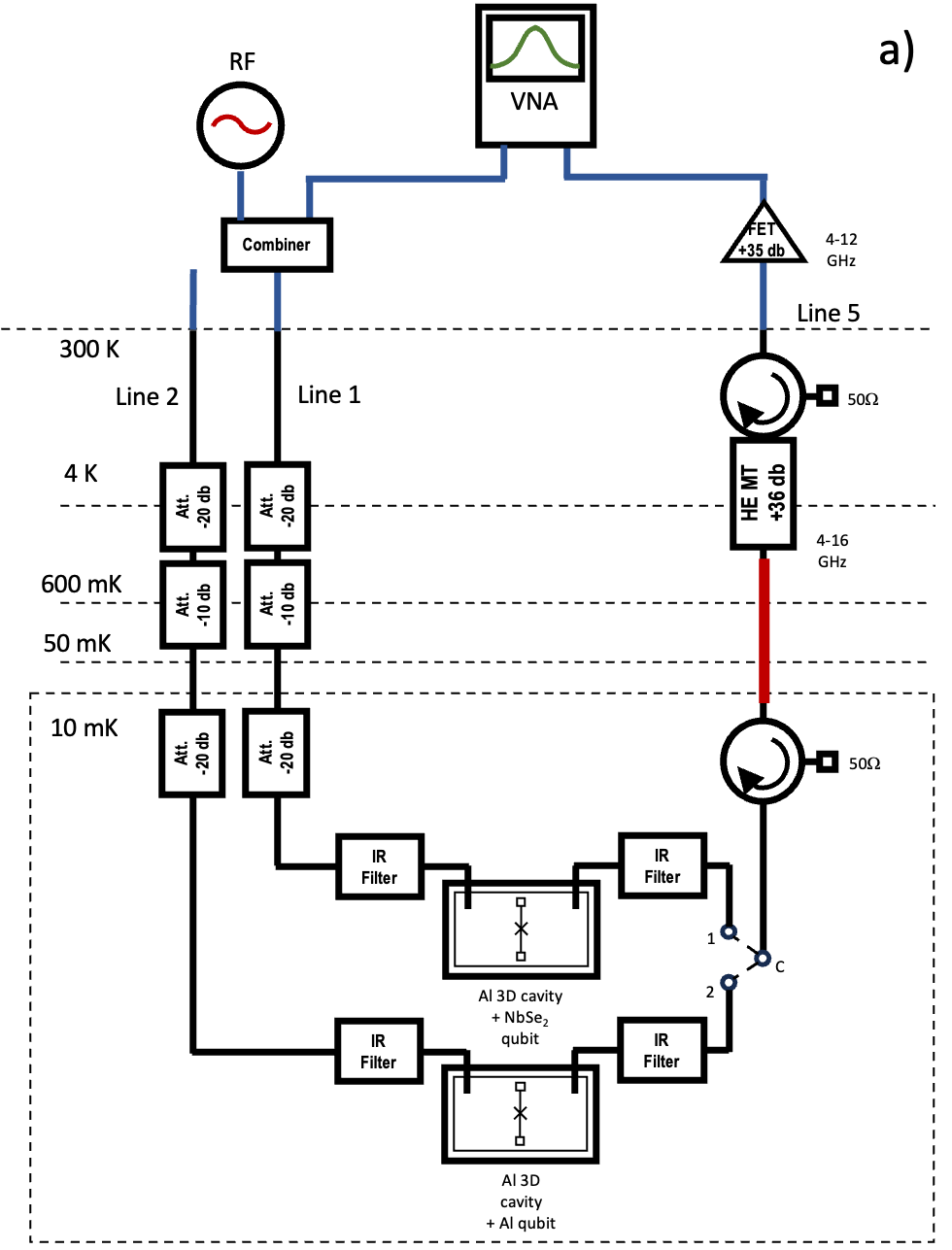}
        \phantomcaption 
        \label{2cavityscheme}
    \end{subfigure}
    
    \vspace{0.5em} 

    \begin{subfigure}{0.48\textwidth}
        \includegraphics[width=\textwidth]{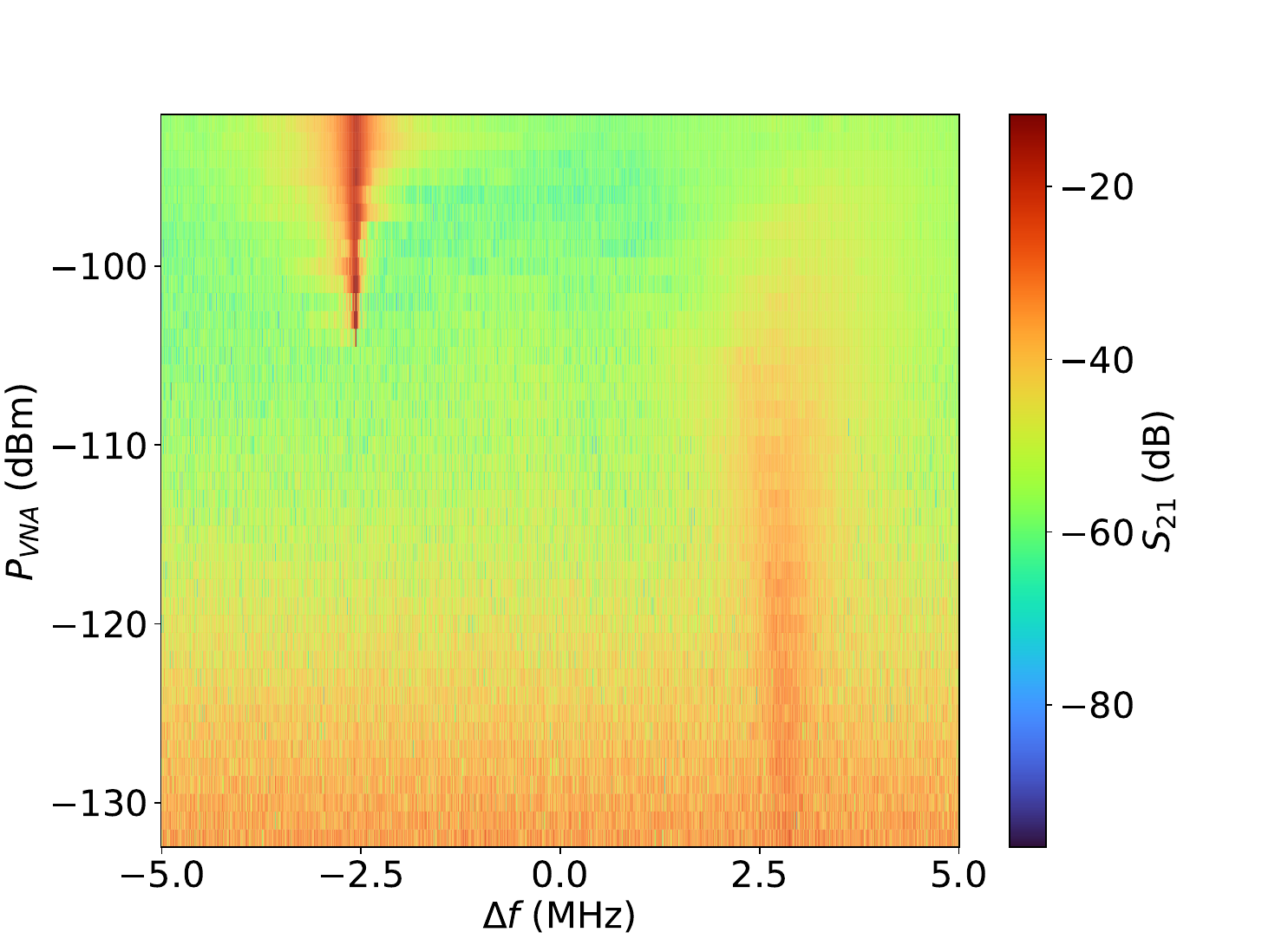}
        \phantomcaption 
        \label{nocirc}
    \end{subfigure}
    \hfill
    \begin{subfigure}{0.48\textwidth}
        \includegraphics[width=\textwidth]{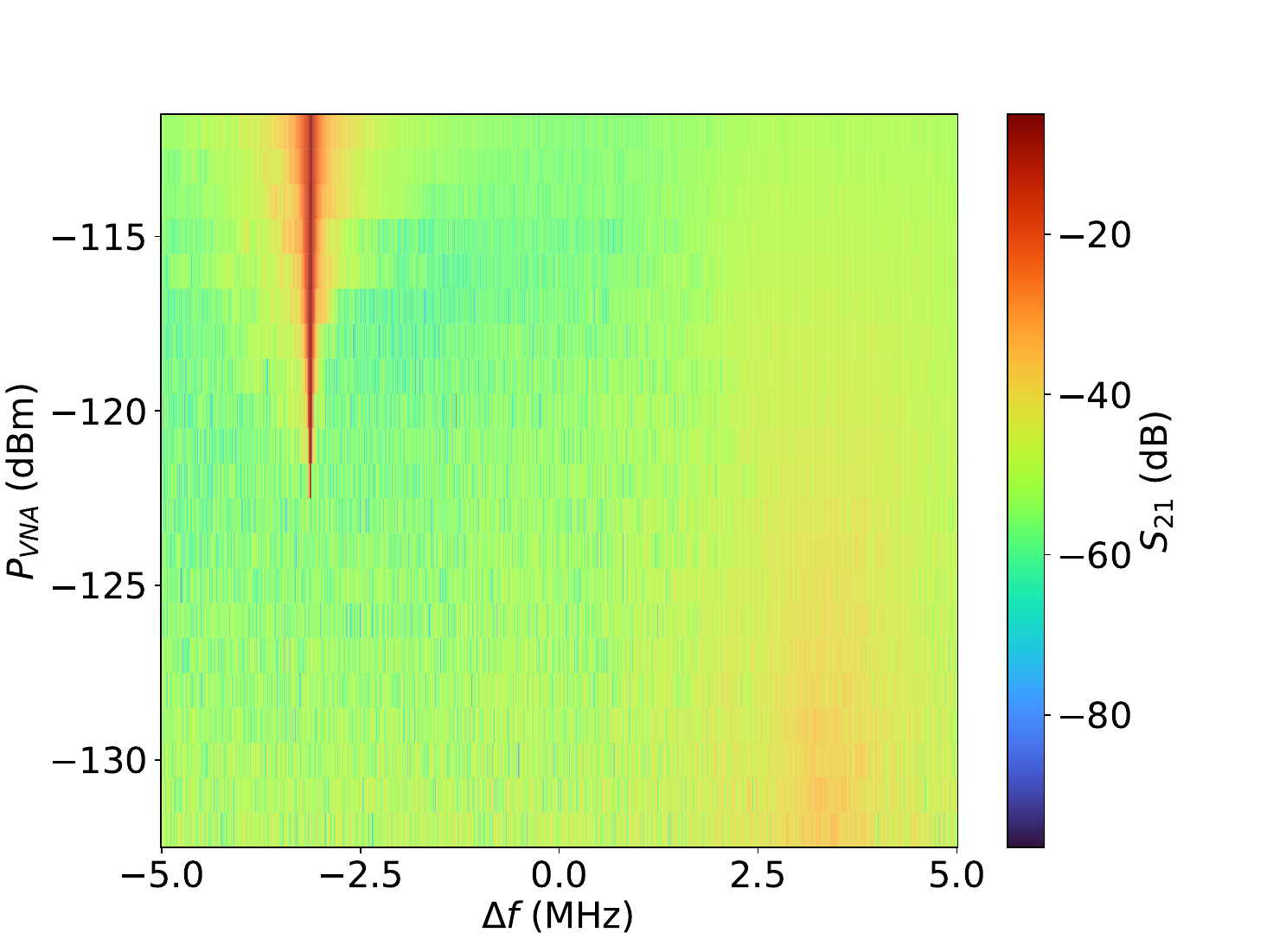}
        \phantomcaption 
        \label{sicirc}
    \end{subfigure}

    \caption{a) RF scheme of the experimental set-up with two cavities and two qubits.  Resonator spectroscopy of the Al qubit measured with  the 10~mK isolator (b) and without  the 10~mK isolator (c). The x axis is expressed as detuning from $f=7.2553$ GHz.}
    \label{fig:circulators}
\end{figure}

In our experimental setup we had a second qubit, an Al transmon hosted in its Al cavity. This qubit was placed in the cryostat together with the NbSe$_2$ device and was connected to the same input and output RF-lines  (See Fig.~\ref{2cavityscheme}). The cavity of the Al qubit was geometrically identical to the NbSe$_2$ one. We performed two different experimental runs, with and without the isolator on the 10~mK stage. This allowed us to verify the photon noise resistance of the NbSe$_2$ qubit. 

\begin{figure}
    \centering
    
    \begin{subfigure}{0.48\textwidth}
        \includegraphics[width=\textwidth]{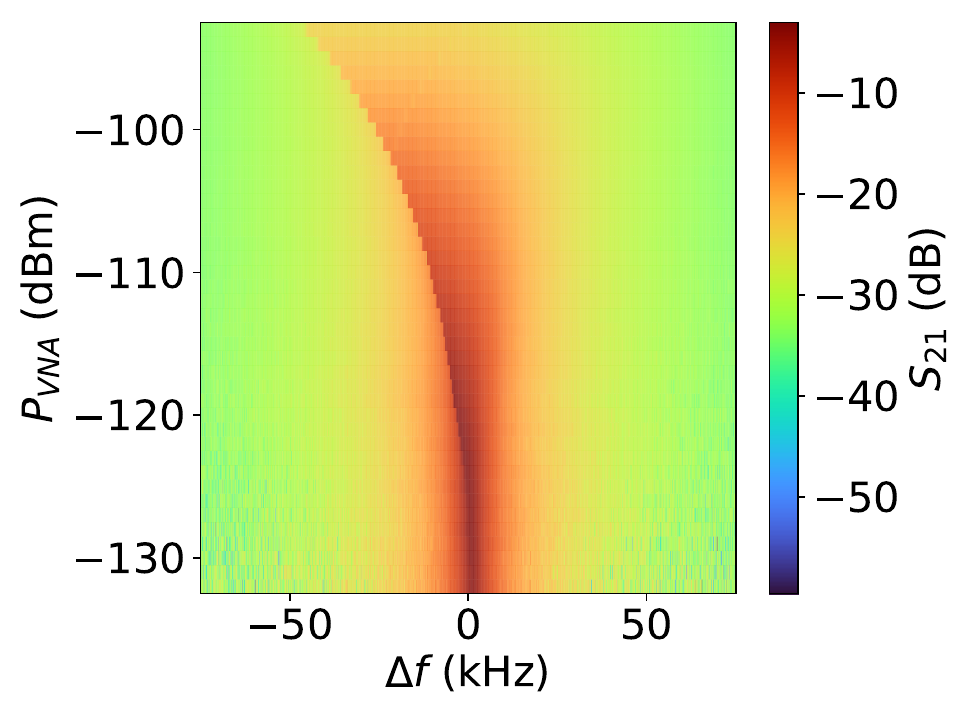}
        \phantomcaption 
        \label{NbSe2nocirc}
    \end{subfigure}
    \hfill
    \begin{subfigure}{0.48\textwidth}
        \includegraphics[width=\textwidth]{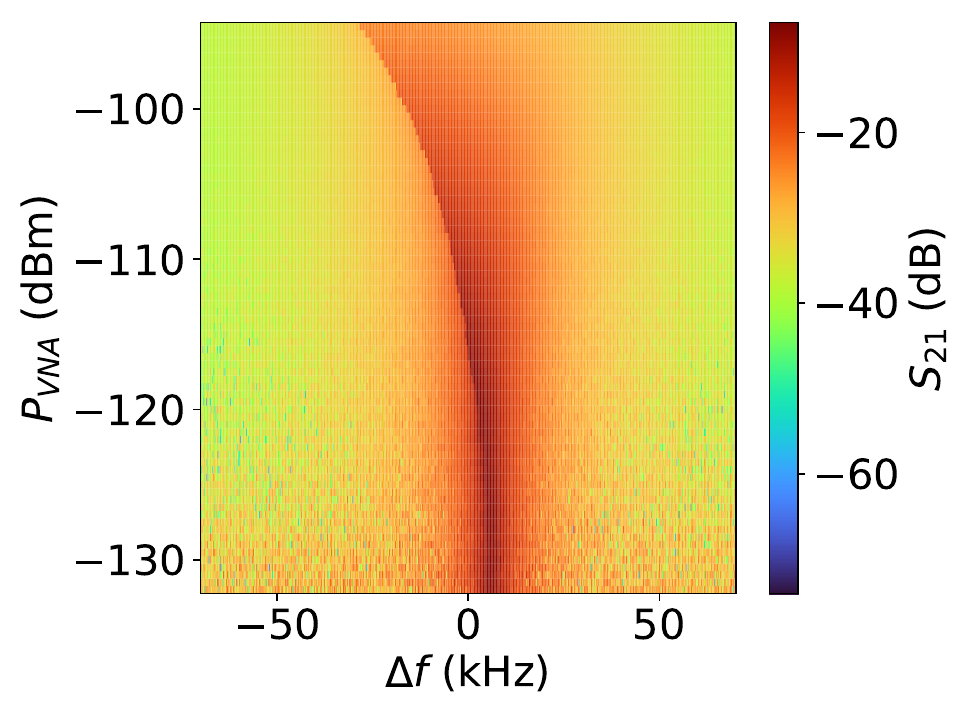}
        \phantomcaption 
        \label{NbSe2sicirc}
    \end{subfigure}

    \caption{Resonator spectroscopy of the NbSe$_2$ qubit measured with  the 10~mK isolator (a) and without the 10~mK isolator (b).The x axis si reported as detuning from f=7.1873 GHz.}
    \label{fig:NbSe2circulators}
\end{figure}

In absence of the 10~mK circulator, the devices were not screened by the photon noise coming from the higher temperature stage of the cryostat, and experienced the emission of a 4~K black-body radiation coming from the HEMT. In this case, we estimated the temperature inside the cavity by assuming that the input and output ports had the same coupling $\gamma$ to the cavity. The input power, dominated by the 4~K radiation, is  $P_{in}=4 k_B \gamma$  while the total power emitted from the two ports is $P_{out}=k_B T_{cav}2\gamma$, where $T_{cav}$ is the cavity temperature. At thermal equilibrium, and neglecting losses in the cavity and in the coax cables, $P_{in}=P_{out}$ and $T_{cav}\simeq2$~K, equivalent to $\sim$ 5 thermal photons on average. The effect of the thermal noise for the Al device is evident in Fig.~\ref{nocirc} and~\ref{sicirc} where we compare the distributions of the cavity transmission as a function of power and frequency of the excitation tone in the two experimental configurations. The power for which the bare resonator frequency appears increases from -122~dBm without isolator to -104~dBm with isolator.
In Fig.~\ref{fig:NbSe2circulators} we report the resonator spectroscopy measurements for the NbSe$_2$ qubit acquired in the same conditions as those reported in Fig.~\ref{fig:circulators}. The effect of the insulator is minimal confirming the high photon noise resistance of our device. 

The high resistance to photon noise of the NbSe$_2$ qubit is due to its high critical current $I_c$. JJs have, in fact, a transition to the normal state when they are crossed by a current larger than $I_c$. This happens when the inductive energy $E_L=L_J I_c^2/2=E_J/2$, where $L_J=\phi_0/(2\pi I_c)$ is the junction inductance, $\phi_0$ the flux quantum and $E_J=\phi_0I_c/2\pi$ the Josephson energy. Then, when the qubit energy is $\hbar\omega_q n_q\simeq E_J$, the junction goes to the normal state and we observe the bare cavity frequency. Driving resonantly a critically coupled cavity with power $P_{in}=\hbar\omega_r \langle a^{\dagger}a\rangle \gamma$ we expect the qubit to have an average occupancy number
\begin{equation}
n_b=\langle b^{\dagger}b\rangle=\left(\frac{g}{\Delta}\right)^2 \langle a^{\dagger}a\rangle=\left(\frac{g}{\Delta}\right)^2 n_a\,,
\end{equation}
where $\Delta=\omega_r-\omega_q$ is the qubit-cavity detuning and $g$ their coupling. Combining these relations with the condition for the transition of the junction to the normal state, we derive the critical value of power $P_{in}$ for which this happens:
\begin{equation}
\label{eq:criticalpower}
    P_{in}^{\star} = \frac{\phi_0I_c}{2\pi}\frac{\omega_r}{\omega_q}\left(\frac{\Delta}{g}\right)^2\gamma\,.
\end{equation}
For the Al transmon with $I_c=20$~nA, $\omega_q/2\pi=5.7$~GHz, $g/2\pi=90$~MHz, $\omega_r/2\pi=7.5$~GHz, $\gamma/2\pi\sim 200$~kHz, we obtain $P_{in}^{\star}=-100~\mbox{dBm}$ in agreement with the value observed in Fig.~\ref{sicirc}. For the NbSe$_2$ junction, with $I_c\simeq30~\mu$m, we then expect a critical power three orders of magnitude larger.

\section{Sign inversion of the of the cavity self-Kerr term 
}\label{pinneswap}
To understand the origin the of the sign inversion of the cavity self-Kerr term when the qubit is excited,
we consider the Hamiltonian of a harmonic resonator coupled to an anharmonic one~\cite{blais2021circuit}
\begin{eqnarray}
H_0&=& \hbar \omega_{r}a^{\dag}a+\hbar \omega_{q}b^{\dag}b
\\\nonumber
H_{NL}&=& -E_{J}\left[ \cos{\phi_b} + \frac{\phi_b^2}{2} \right]
\\\nonumber
H_{Int}&=& \hbar g\left(a b^{\dag}+a^{\dag}b \right)\,,
\end{eqnarray}
where
\begin{equation}
\phi_{b}=\phi^{ZPF} (b^{\dag}+b)
\end{equation}
with
\begin{equation}
\phi^{ZPF}=\left(\frac{2 E_C}{E_J}\right)^{1/4}
\end{equation}
and $E_C=e^2/2C$.
By diagonalizing the dipole interaction term $H_{Int}$ through the unitary Bogoliubov transformation
\begin{equation}
 U=exp\left[\Lambda(a^{\dag}b-ab^{\dag}) \right]
\end{equation}
expanding for small $\Lambda\simeq g/\Delta$ and keeping terms up to sixth order in the annihilation and creation operators,  we find the cavity Kerr-term 
\begin{equation}
\label{eq:ker2}
    H_{Kerr}=\frac{K_a}{2}\left[ 1- \frac{1}{2}\phi_b^2\right] a^{\dag 2}a^2\,.
\end{equation}
where $K_a$ is defined in Eq.~\ref{selfcavity}.
In Eq.~\ref{eq:ker2} when $\phi_b^2\sim b^{\dag}b$ is large enough, the sign of the non linear coefficient is switched from negative to positive. 
However, this calculation is only valid in the small power regime and provide a qualitative understanding of the effect.

\section{Determination of the qubit coupling to the readout mode TM110}
\label{cavityvsdrivepower}
\begin{figure}[h]
    \centering
    \includegraphics[width=0.5\linewidth]{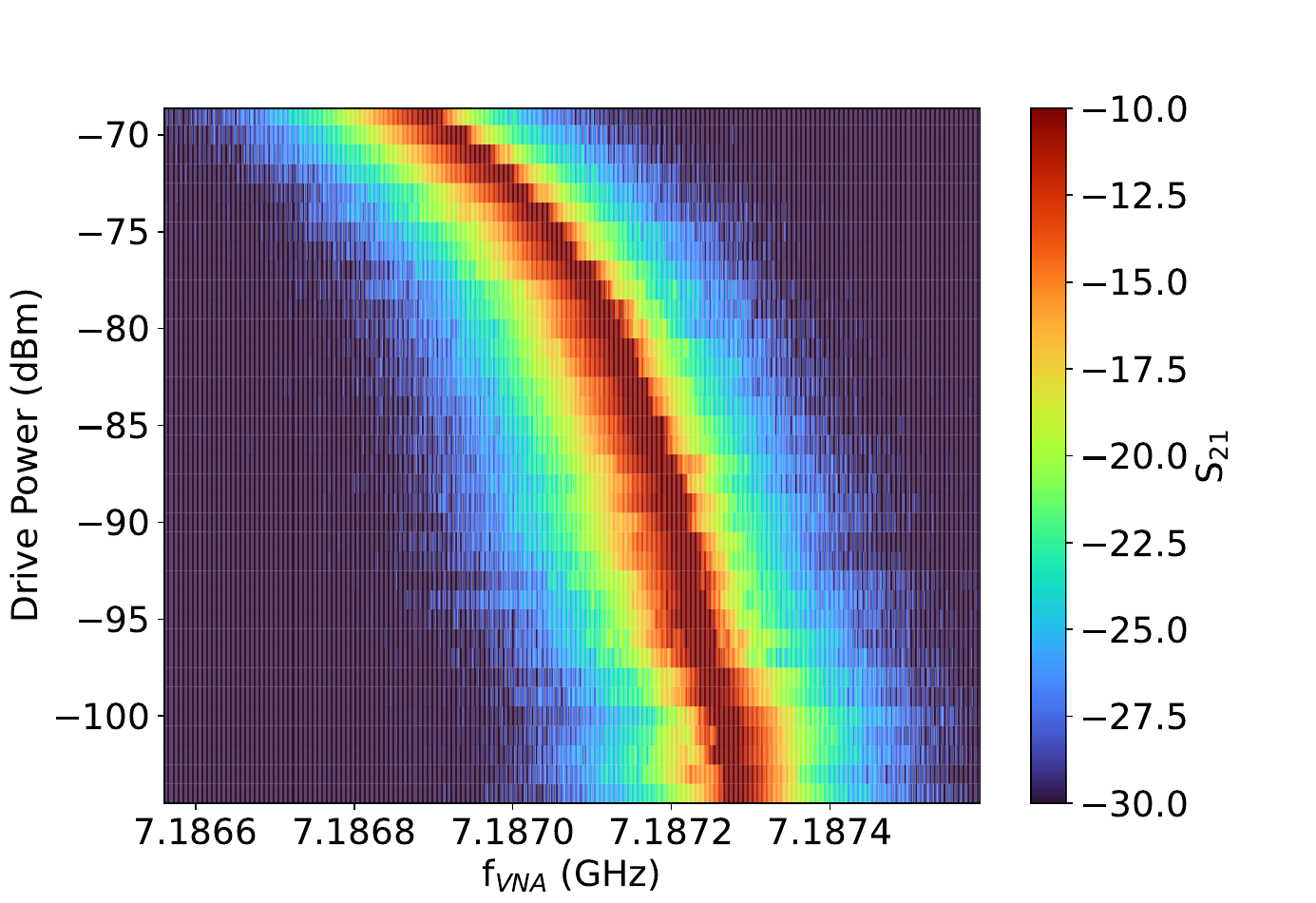}
    \caption{Dressed cavity transmission as a function of the qubit drive excitation power}
    \label{fig:vna_vs_drive}
\end{figure}

We estimated the dipole coupling $g_{110}/2\pi$ to the cavity mode $TM_{110}$ by measuring the dressed-cavity frequency while driving the qubit at different powers. The drive tone was set in resonance with the qubit at the frequency $\omega_d =12.611$~GHz. The value of the cavity frequency was measured from the cavity transmission at different drive-powers as shown in Fig.~\ref{fig:vna_vs_drive}. We used the readout power P$=-121$~dBm at the cavity port while varying the qubit drive-power in the range $(-104; -69)$~dBm similar to the one used for the two-tone spectroscopy (See Fig.~\ref{fig:2tones}). This allowed us to plot the dressed-cavity frequency as a function of the qubit frequency (See Fig.~\ref{fig:vwvsvr}) and derive their relation by combining Eq.~\ref{vr vs vq-1} and Eq.~\ref{vr vs vq-2} into Eq.~\ref{vr vs vq-3}.

\section{Determination of the qubit coupling to the mode TM130 by Self-Kerr effect} \label{kerrdata}

\begin{figure}[h!]
    \centering
    
    \includegraphics[width=0.6\textwidth]{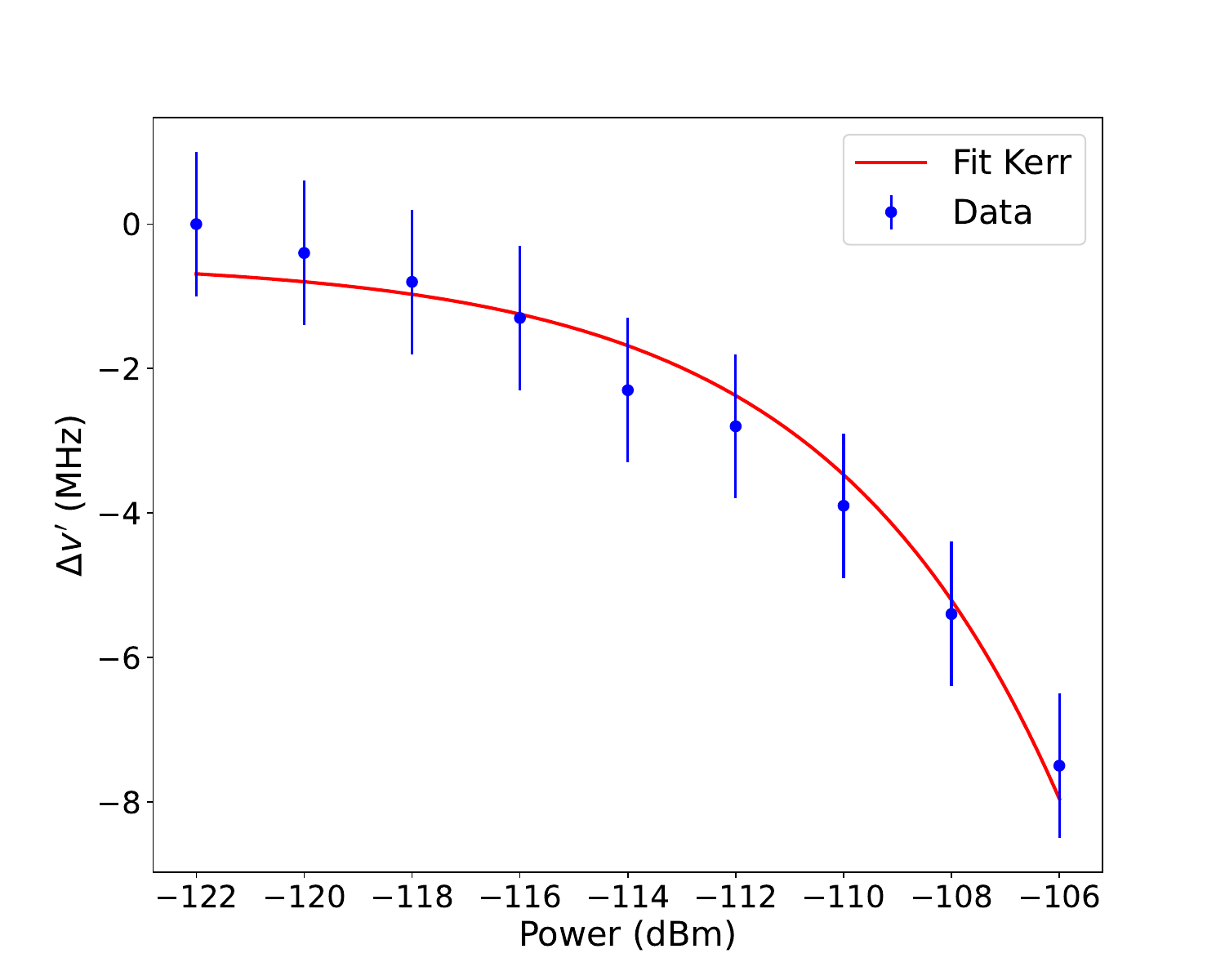}
    \caption{Dressed cavity resonance frequency (expressed as detuning from $\nu_c=13.4848$~GHz) as a function of the readout power.}
    \label{fig:selfcavitykerr}
\end{figure}

We estimated the coupling between the qubit and the mode TM$_{130}$ first by
determining the cavity self-Kerr term from the measurement of the frequency shift of the cavity as a function of the readout-tone power and then calculating the coupling $g_{130}/2\pi$ from Eq.~\ref{selfcavity}. Data are reported in Fig.~\ref{fig:selfcavitykerr}.

We interpolated the experimental data using:
\begin{equation}
    \omega_r^{\prime}=\omega_r+\frac{1}{2}K_a n
    \label{kerrfit}
\end{equation}
where $\omega_r^{\prime}$ is the dressed-cavity frequency, $K_a$ the cavity self-Kerr and $n$ the number of photons inside a critically coupled cavity:  
\begin{equation}
    n=\frac{P}{\gamma \hbar\omega_r}
\end{equation}
where $P$ is the power of the excitation tone, $\omega_r/2\pi$ the cavity frequency and $\gamma$ the cavity linewidth. The power $P$ is computed taking into account the attenuation of the RF lines equal to $-85$~dB and using the values $\gamma/2\pi=200$~kHz and $\alpha/2\pi=-1.3$~MHz obtain from simulation.
The fit to the data points returned $K_a= 13.33$~kHz. Inverting Eq.~\ref{selfcavity} we calculated $g_{130}/2\pi=257\pm60_{syst}$~MHz which is in reasonable agreement  with the value of the EPR analysis. We estimated the systematic error by considering an order of magnitude variation on cavity coupling and qubit anharmonicity.

\section{Rabi frequency of an off-resonatly driven qubit} \label{cavityattenuationII}

We consider here the Hamiltonian for a qubit-cavity system driven from port 1 at a frequency $\omega_d$ close to the qubit frequency:
\begin{eqnarray}
H&=&\omega_r a^{\dagger}a+\omega_q b^{\dagger}b+\frac{\alpha}{2}b^{\dagger 2}b^2
\\\nonumber
&&+ g(b^{\dagger}a+ba^{\dagger})+\gamma_1 \left( \epsilon(t) a^{\dagger}+\epsilon(t)^* a\right)\,.
\end{eqnarray}
The average occupancy number in the resonator is:
\begin{equation}
\label{eq:cavityoccupation}
|\epsilon|^2=\langle a^{\dag}a\rangle=\frac{P_{in}}{\hbar\omega_r}\frac{4\gamma_1}{\gamma_{tot}^2}
\end{equation}
where $\gamma_{tot}=\gamma_1+\gamma_2+\gamma_0$ with $\gamma_0$ the internal losses and $\gamma_{1,2}$ the width of ports 1 and 2. 
Performing the time-dependent unitary transformation $U_p=\exp(\xi(t)^* a-\xi(t) a^{\dagger})$ the Hamiltonian transforms as
\begin{equation}
   H^{\prime} = U^{\dagger}_p H U_p -i U^{\dagger}_p \dot{U}_p
\end{equation}
and, using the following relations
\begin{eqnarray}
U^{\dagger}_p a U_p &=& a - \xi
\\\nonumber
U^{\dagger}_p \dot{U}_p &=& \dot{\xi}^*a-\dot{\xi}a^{\dagger}
+\frac{1}{2}\left(  \dot{\xi}\xi^*-\dot{\xi}^*\xi \right)\,,
\end{eqnarray}
it becomes
\begin{eqnarray}
\label{eq:app-qubitdrive}
H&=&\omega_r a^{\dagger}a+\omega_q b^{\dagger}b+\frac{\alpha}{2}b^{\dagger 2}b^2
\\\nonumber
&&+ g(b^{\dagger}a+ba^{\dagger})-g \left( \xi(t) b^{\dagger}+\xi(t)^* b\right)
\label{drive_hamiltonian}
\end{eqnarray}
where to cancel the cavity-drive interaction term we chose
\begin{equation}
\dot{\xi}=-i\omega_r\xi+i\gamma_1 \epsilon_p(t)\,.
\end{equation}
For a harmonic drive $\epsilon(t)=\epsilon\exp(-i\omega_d t)$ 
\begin{equation}
\xi=-\frac{\gamma_1\epsilon}{\Delta}e^{-i\omega_d t}
\end{equation}
with $\Delta=\omega_d-\omega_r$.

Finally, the qubit-drive interaction terms is
\begin{equation}
H_{qp}=-g \frac{\gamma_1\epsilon}{\Delta}\left( e^{-i\omega_d t} b^{\dagger}+e^{i\omega_d t} b\right)
\end{equation}
corresponding to a Rabi frequency
\begin{equation}
\omega_{Rabi}=2 g \sqrt{\frac{\gamma_1 P_{in}}{\Delta^2 \hbar\omega_d}}=2g\sqrt{n_{eff}}\,.
\end{equation}
where, the effective number of photons in the cavity is:
\begin{equation}
n_{eff}=\frac{\gamma_1\epsilon}{\Delta}=\frac{\gamma_1}{\Delta}\sqrt{\frac{P_{in}}{\hbar\omega_r}\frac{4\gamma_1}{\gamma_{tot}^2}}\,.
\end{equation}
Qubit losses slightly shifts the Rabi frequency to~\cite{Petruccione}
\begin{equation}
    \Omega_{Rabi}^{'}=\sqrt{\omega_{Rabi}^2-\left(\frac{\Gamma_{q}}{4} \right)^2}\,.
\end{equation}
To simulate the  Rabi oscillation of the qubit as a function of the drive amplitude we considered the closest mode TM$_{130}$ at frequency 13.45~GHz. Adopting the RWA and considering only the qubit-drive system  the Hamiltonian is
\begin{equation}
    H=\frac{K_b}{2}b^{\dagger2}b^2+g_{130}\sqrt{n_{eff}}(b^{\dagger}+b)\,.
    \label{simulatio_hamil}
\end{equation}

\begin{figure}[h!]
    \centering
    
    \includegraphics[width=0.6\textwidth]{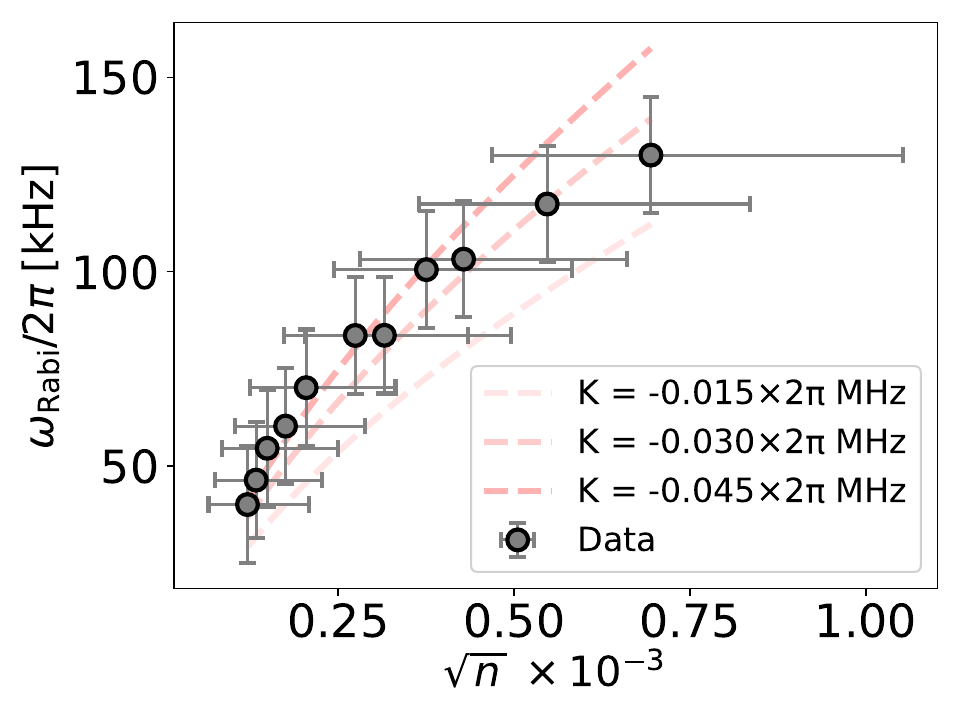}
    \caption{Rabi frequency as a function of the drive amplitude expressed as squared root of the effective
cavity occupation-number. Along with the experimental data are reported the simulations obtained by solving the time evolution of Eq.~\ref{simulatio_hamil} for different values of K with $g_{130}/2\pi=210$~MHz.}
    \label{fig:rabisimul_vs_alpha}
\end{figure}

We reproduced the trend of the Rabi frequency as a function of the drive power shown in Fig.~\ref{fig:rabisimul_vs_alpha} by setting $g_{130}/2\pi=210\pm60$~MHz and $K_b/2\pi=\alpha/2\pi=-0.03\pm0.015$~MHz, considering the total width of the mode TM$_{130}$  $\gamma_{tot}=200$~kHz as derived from the simulations and assuming an under-critical coupling of the input port $\gamma_1=10$~kHz. We considered the qubit relaxation-rate $\Gamma_{q}=0.153$~MHz and a dephasing rate  $\Gamma_{\Phi}=\Gamma_{q}/2$. Finally, we considered an attenuation in the input line of -95~dB adding extra 10~dB with respect to our estimate, motivated by the large uncertainty due to strong reflections at these frequencies. We estimated the systematic uncertainties on $g_{130}$ and $\alpha$ varying their value up to the point where the output simulation was not compatible with the data error bars as reported in Fig.~\ref{fig:rabisimul_vs_alpha} and \ref{fig:rabivspower}

\end{document}